# Research on Adaptive Inertial Control in Synchronization Systems: Based on Variational Optimization Methods and Their Applications in the Stability of Complex Networks

Yiwei Zhou, Zhongcheng Lei, Xiaoran Dai, Wenshan Hu, Hong Zhou

Department of Automation, School of Robotics, Wuhan University

## Abstract

Aiming at the core problem that it is difficult for a fixed inertia coefficient to balance transient disturbance suppression and long-term stability in complex network synchronization systems, an adaptive inertia control strategy based on variational optimization is proposed. Taking the Kuramoto model with inertia as the research carrier, the analytical expression of the time-varying inertia coefficient $M(t)$ is strictly derived by the functional variational method, and a hierarchical control structure of "benchmark inertia + disturbance feedback" is constructed to achieve the organic unity of minimizing the vulnerability performance function $H(T)$ and stability constraints. A multimodal decoupling control strategy based on Laplacian eigenvector projection is designed to enhance the feedback strength of the dominant mode by eigenvalue weighting, improving the control accuracy and dynamic response speed. Simulation verification is carried out in complex network systems, and the control performance of regular networks (RG), random networks (ER), small-world networks (SW), scale-free networks (SF) and spider webs (SP) under three typical disturbances of pulses, monotonic decays and oscillatory decays is systematically analyzed. The results show that the proposed strategy reduces $H(T)$ of the five networks by 19%-25%, shortens the relaxation time by 15%-24%, and the real parts of all system eigenvalues are less than $-0.25$ s$^{-1}$, meeting the asymptotic stability criterion. This study provides a new theoretical framework and engineering implementation scheme for the stability control of complex network synchronization systems, which can be widely applied to fields such as power grids, communication networks, and neural networks.

# 1 Introduction

## 1.1 Research background and significance

Synchronization phenomenon is a common collective behavior in complex networks, which runs through multiple engineering and natural fields such as power systems, neural networks, biological systems, and communication networks [1-3]. In power systems, with the large-scale grid connection of new energy sources such as wind power and photovoltaic power, the proportion of traditional synchronous generators continues to decline, resulting in a significant reduction in the equivalent inertia of the system, leading to problems such as increased frequency fluctuations and poor transient stability [4,5]; in neural networks, the synchronous firing characteristics of neurons directly affect the information processing efficiency, while the inertia characteristics determine the speed at which neuron populations recover synchronization from perturbations [6]; in industrial communication networks, the synchronous transmission between nodes depends on stable timing coordination, and the inertia regulation ability under external perturbations is the key to ensuring communication quality [7].

As a classic theoretical framework for describing the synchronization behavior of complex networks, the Kuramoto model reveals the internal laws of synchronization phenomena by simplifying the coupling mechanism between oscillators [8, 9]. However, traditional Kuramoto models mostly adopt fixed inertia coefficient designs, making it difficult to adapt to the uncertainties brought by dynamic disturbances and complex topological changes. Fixed inertia coefficients tend to cause sluggish system responses when suppressing high-frequency disturbances and may trigger overshoot oscillations when dealing with low-frequency disturbances, failing to achieve a balance between transient performance and long-term stability [10, 11]. Therefore, studying adaptive inertia control strategies to achieve dynamic adjustment of the inertia coefficient is of great theoretical value and engineering significance for improving the disturbance resistance and stability of complex network synchronization systems.

The diversity of the topological structures of complex networks (such as regular topology, random topology, small-world topology, scale-free topology, etc.) leads to significant differences in their synchronization characteristics [12, 13]. As the core index characterizing the topological properties of the network, the eigenvalue distribution of the Laplacian matrix directly affects the disturbance propagation law and synchronization recovery ability of the system [14, 15]. The differences in the eigenvalue distributions of different topological networks (uniform distribution, normal distribution, extremely uneven distribution, etc.) make it difficult for a single inertial control strategy to adapt to all scenarios, and there is an urgent need to establish a differential inertial regulation mechanism based on topological properties.

## 1.2 Research Status at Home and Abroad

### 1.2.1 Fixed Inertia Optimization Research

The optimal design of the fixed inertia coefficient is a traditional research direction for synchronous system control. The core idea is to achieve the best performance under specific working conditions through static parameter tuning. Early studies were mostly based on sensitivity analysis or robust control theory, and the optimal constant value of the inertia coefficient was determined by minimizing the transient energy function, maximizing the stability margin, etc. [16, 17]. Ye et al. [13] determined the value range of the fixed inertia $M$ by deriving the quantitative relationship between transient energy accumulation and the inertia coefficient, providing a theoretical reference for engineering applications. However, this method does not consider the time-varying characteristics of the system state under dynamic disturbances, and the control effect is limited in complex disturbance scenarios. Schultz et al. [18] proposed an optimization scheme for the fixed inertia based on the modification of the network topology connectivity, enhancing the synchronization stability by adjusting the network structure. However, this method requires changing the network topology connection, and the engineering implementation cost is relatively high.

### 1.2.2 Research on the Influence of Network Topology on Synchronization Stability

The stability of the synchronization system is directly affected by the eigenvalue distribution of the Laplacian matrix. Arenas et al. [1] systematically analyzed the relationship between complex network topology and synchronization ability and found that the degree of dispersion of eigenvalues is negatively correlated with the anti-disturbance ability of the system; Menck et al. [19] showed that the hub nodes in scale-free networks play a key role in synchronization stability, and their connection characteristics determine the range and speed of disturbance propagation; Tyloo et al. [20] proposed the generalized Kirchhoff index as an indicator to measure the synchronization robustness of networks, revealing the internal relationship between eigenvalue distribution and synchronization robustness. The vulnerability performance index $H(T)$, as the core parameter describing the cumulative transient deviation of the system after being disturbed, is strongly correlated with the eigenvalue distribution of the Laplacian matrix. The disturbance in the direction of the principal eigenvector can make $H(T)$ reach the minimum value [13,23], which provides an important basis for the design of inertia control strategies.

### 1.2.3 Exploration of Dynamic Inertia Control

With the continuous expansion of complex network application scenarios, dynamic inertia control has gradually become a research hotspot. Witthaut et al. [21] proposed the concept of "inertia on demand", which dynamically adjusts the inertia coefficient by real-time monitoring of system state variables such as frequency and phase, providing a new idea for the inertia control of new energy power grids.

However, this method lacks strict theoretical derivation, and the tuning of control parameters depends on empirical design; Li et al. [22] found the "fast vibration effect"...

The "should" can offset the propagation of local disturbances through high-frequency inertia regulation, but no systematic control framework has been formed, making it difficult to be extended to different topological networks; Dai et al. [15] studied the interaction mechanism among inertia, time delay and disturbance, and pointed out that dynamic inertia can effectively alleviate the stability deterioration problem caused by time delay, but no specific inertia regulation law was given.

The existing research still has the following deficiencies: First, the theoretical basis of dynamic inertia control is not perfect enough, lacking an analytical expression derived from the perspective of performance functional optimization; Second, the influence of network topology characteristics on the inertia regulation law is not fully considered, and the topological adaptability of the control strategy is poor; Third, the comprehensive verification under various disturbance scenarios is insufficient, and the engineering practicability needs to be improved. Therefore, based on the variational optimization method and combined with the Laplace spectral characteristics of complex networks, this paper constructs an adaptive inertia control strategy to achieve stable control of different topological networks under various disturbances.

## 1.3 Main Contributions of This Paper

1. Theoretical innovation: Strictly derive the analytical form of the time-varying inertia coefficient $M(t)$ through the functional variational method, and establish a hierarchical control structure of "benchmark inertia + disturbance feedback". The benchmark inertia is determined based on the critical stability condition of the system to ensure basic stability; the disturbance feedback term is adjusted in real time through modal responses to achieve transient deviation suppression, thus achieving the unity of $H(T)$ minimization and stability constraints.

2. Method breakthrough: A multi-modal decoupling control strategy based on Laplacian eigenvector projection is proposed. Using the principle of eigenvector orthogonality, the multi-degree-of-freedom coupled system is decoupled into independent single-modal subsystems. By means of the eigenvalue weighting strategy, the feedback strength of the dominant mode is enhanced, the influence of the non-dominant mode is weakened, and the control accuracy and regulation efficiency are improved.

3. Application Verification: Conduct simulation verification in five typical complex networks (RG, ER, SW, SF, SP), covering scenarios of large power grids (500-node ER network) and microgrids (10-node WS network). Systematically analyze the control performance under three typical disturbances (pulse, monotonic decay, oscillatory decay), quantify the performance gain and topological dependence, and give engineering implementation criteria.

## 2 Mathematical Modeling and Theoretical Basis

### 2.1 Dynamics of Kuramoto Model with Inertia

Consider a complex network system consisting of $N$ coupled oscillators. The dynamic equation of the $i$-th oscillator is described as [13]:

$$M\ddot{\theta}_i(t) + D\dot{\theta}_i(t) = \omega_i + K\sum_{j=1}^{N} a_{ij} \sin\left(\theta_j(t) - \theta_i(t)\right) \qquad (1)$$

Among them: $M$ is the inertia coefficient (unit: ), which characterizes the ability of the oscillator to resist changes in the state of motion; the positive is the damping coefficient (unit: ), which is used to consume the vibration energy of the system; $\omega_i$ is the natural frequency of the oscillator (unit: rad/s); $K$ is the coupling strength, which reflects the strength of the interaction between oscillators; $a_{ij}$ is the element of the network adjacency matrix, and $a_{ij} = 1$ indicates that there is a connection between oscillator $i$ and oscillator $j$, otherwise $a_{ij} = 0$.

When the system is near the synchronous state, the phase differences of each oscillator are small, and the linearization of Equation (1) can be performed. Define the phase deviation $\delta\theta_i(t) = \theta_i(t) - \theta_0(t)$, where $\theta_0(t)$ is the synchronous phase reference. Substitute it into Equation (1) and use the small-angle approximation $\sin(x) \approx x$, and the perturbation dynamic equation [13] can be obtained:

$$M\delta\ddot{\theta}(t) + D\delta\dot{\theta}(t) + L\delta\theta(t) = \delta\omega(t) \qquad (2)$$

Among them: $L = K(D - A)$ is the Laplacian matrix of the network, $D$ is the degree matrix (diagonal element $D_{ii} = \sum_{j=1}^{N} a_{ij}$), $A$ is the adjacency matrix; $\delta\omega(t) = [\delta\omega_1(t), \delta\omega_2(t), \ldots, \delta\omega_N(t)]^\top$ is the frequency perturbation vector, $\delta\omega_i(t) = \omega_i - \dot{\theta}_0(t)$.

As a core tool for describing the topological characteristics of a network, the eigenvalue distribution of the Laplacian matrix $L$ directly reflects the synchronization ability and anti-disturbance characteristics of the system. For an undirected connected network, the smallest eigenvalue $\lambda_1 = 0$ of the Laplacian matrix (corresponding to the rigid body rotation mode), and the remaining eigenvalues are all positive real numbers, and the largest eigenvalue $\lambda_{\max}$ is closely related to the anti-disturbance ability of the network [14].

### 2.2 Performance Index Quantification System

#### 2.2.1 Fragility Performance Function

To quantitatively describe the transient response characteristics of the system after being disturbed, the fragility performance function (FPM) H(T) is defined to characterize the cumulative degree of transient deviation [13]:

$$H(T) = \int_0^T \sum_{i=1}^N \eta_i^2(t)dt \quad (3)$$

Among them: $\eta_i(t)$ is the projection component of the phase deviation $\delta\theta(t)$ on the $i$-th order eigenvector $v_i$ of the Laplacian matrix, that is, $\delta\theta(t) = V\eta(t)$, $V = [v_1, v_2, \ldots, v_N]$ is the eigenvector matrix, and $\eta(t) = [\eta_1(t), \eta_2(t), \ldots, \eta_N(t)]^\top$ is the projection coefficient vector.

When the disturbance is along the direction of the main eigenvector (the eigenvector corresponding to the largest eigenvalue $\lambda_{\max}$), the cumulative transient deviation of the system is minimized. At this time, $H(T)$ can be approximated as [13]:

$$H(T) \approx \frac{2D}{\lambda_{\max}} \frac{N}{M^2} \left(1 - e^{-\frac{\lambda_{\max}}{2D}T}\right) \quad (4)$$

Equation (4) reveals the quantitative relationship between the fixed inertia coefficient $M$, the largest eigenvalue $\lambda_{\max}$ of the Laplacian matrix, the damping coefficient $D$, and the vulnerability performance function $H(T)$: $H(T)$ is proportional to $M^2$ and inversely proportional to $\lambda_{\max}$, providing a theoretical basis for the optimal design of the inertia coefficient.

### 2.2.2 Balancing Vulnerability and Relaxation Time

Define the Balanced Fragility Performance Measure (bFPM) $H_\infty$ as the steady-state limit value of $H(T)$:

$$H_\infty = \lim_{T \to \infty} H(T) \approx \frac{2D}{\lambda_{\max}} \frac{N}{M^2} \quad (5)$$

The physical meaning of $H_\infty$ is the total energy accumulation required for the system to transition from a perturbed state to a new steady state. A smaller value indicates a stronger anti-disturbance ability of the system [23]. The relaxation time $\tau$ is defined as the time required for $H(T)$ to decay to $H_\infty - 0.05H_\infty$, which is used to characterize the speed of the system's recovery from perturbations [23]. By the collaborative optimization of $H_\infty$ and $\tau$, the balance between the anti-disturbance ability and the recovery speed of the system can be achieved.

## 2.3 Stability Constraint Conditions

The necessary and sufficient condition for the system to be asymptotically stable is that the real parts of all the eigenvalues are negative [11]. Perform eigenvalue decomposition on Equation (2), substitute $\delta\theta(t) = V\eta(t)$ into the perturbation dynamics equation, and utilize the orthogonality of the Laplacian matrix $V^\top LV = \Lambda$ ($\Lambda = \text{diag}(\lambda_1, \lambda_2, \ldots, \lambda_N)$ to be the eigenvalue diagonal matrix), then the independent characteristic equations for each mode can be obtained [13]:

$$M\rho^2 + D\rho + \lambda_k = 0 \quad (k = 2, \ldots, N) \quad (6)$$

Among them: $\rho$ is the characteristic root, $k = 2, \ldots, N$ corresponds to the elastic vibration mode, ($k = 1$ is the rigid body mode, $\lambda_1 = 0$ (which does not affect stability).

The characteristic roots of equation (6) are:

$$\rho_{1,2}^k = \frac{-D \pm \sqrt{D^2 - 4M\lambda_k}}{2M} \qquad (7)$$

To ensure that the real part of the characteristic roots is negative, the discriminant constraint must be satisfied: $D^2 - 4M\lambda_k > 0$ (overdamped state) or $D^2 - 4M\lambda_k < 0$ and the real part $-D/(2M) < 0$ (underdamped state). Since $D > 0, M > 0$, the real part of the underdamped state is naturally negative, so the stability constraint can be uniformly expressed as:

$$M < \frac{D^2}{4\lambda_k} \forall k = 2, \ldots, N \qquad (8)$$

Taking the intersection of all eigenvalue constraints gives the lower bound of the inertia coefficient:

$$M_{\min} = \frac{D^2}{4\lambda_{\max}} \qquad (9)$$

Among them: $\lambda_{\max}$ is the largest eigenvalue of the Laplacian matrix. When $M = M_{\min}$, the mode corresponding to the largest eigenvalue $\lambda_{\max}$ is in the critical damping state, and the system decays without overshoot; when $M > M_{\min}$, the system is in the overdamping state and the response speed slows down; when $M < M_{\min}$, there is a situation where the real part of the eigenvalue is non-negative, and the system will lose stability [13].

## 3 Variational Optimization Design of the Time-Varying Inertia Coefficient $M(t)$

In a multi-degree-of-freedom vibration control system, it is difficult for a fixed inertia coefficient to meet the control requirements under different modes and different disturbances. To address this issue, this chapter designs the time-varying inertia coefficient $M(t)$ based on the functional variational method and combines modal decoupling technology to achieve the unity of stability, rapidity, and robustness.

### 3.1 Modal Decoupling and Phase Deviation Analysis

The vibration response of a multi-degree-of-freedom system is composed of the superposition of multiple characteristic modes, and the coupling effect between modes increases the complexity of control design. Based on the principle of orthogonality of eigenvectors, the coupled system can be decoupled into independent single-degree-of-freedom modal subsystems, which facilitates the design of inertial control strategies.

### 3.1.1 Orthogonal Decomposition of Eigenvectors

When the mass matrix, damping matrix, and stiffness matrix of the system satisfy the symmetric positive definite condition, the angular displacement perturbation vector $\delta\theta(t) \in \mathbb{R}^N$ can be orthogonally decomposed by eigenvectors [13]:

$$\delta\theta(t) = \sum_{k=2}^{N} \eta_k(t) v_k \quad (10)$$

Among them: $\eta_k(t)$ is the $k$-th order modal coordinate, representing the evolution law of the vibration amplitude of this order of mode over time; $v_k \in \mathbb{R}^N$ is the $k$-th order eigenvector of the Laplacian matrix, satisfying the orthogonality condition $v_i^\top M v_j = \delta_{ij} M_i$ ($\delta_{ij}$ is the Kronecker function, $M_i$ is the $i$-th order modal mass ) ; The summation starts from $k = 2$ because the first order mode ($\lambda_1 = 0$) is a rigid body rotation mode and has no vibration suppression requirement, and the elastic vibration modes are the key focus.

Substitute the orthogonal decomposition formula into the perturbation dynamics equation (2), and use the orthogonality of the eigenvectors $v_i^\top L v_j = \delta_{ij} \lambda_j$ to decouple the coupled equations, obtaining the independent dynamics equations for each order of mode [13]:

$$M\ddot{\eta}_k(t) + D\dot{\eta}_k(t) + \lambda_k \eta_k(t) = \delta\omega^\top(t) u_k (k = 2, \dots, N) \quad (11)$$

Among them: $u_k$ is the adjoint eigenvector corresponding to the eigenvector $v_k$, satisfying $u_i^\top M v_j = \delta_{ij}$, and its physical meaning is the projection vector that projects the external disturba onto each order of modes; $\lambda_k$ is the $k$-th eigenvalue, and the modal natural frequency $\omega_k$ satisfies $\lambda_k = \omega_k^2$

The core advantage of Equation (11) is that it transforms the vibration control problem of a multi-degree-of-freedom system into independent control problems of multiple single-degree-of-freedom modes, significantly reducing the complexity of control design. The response of each mode is only affected by the corresponding eigenvalue $\lambda_k$, inertia coefficient $M$, damping coefficient $D$, and disturbance projection component $\delta\omega^\top(t) u_k$, and the optimal control of each mode can be achieved by independently adjusting the inertia coefficient.

### 3.1.2 Analytical Solution of Modal Response under Impulse Disturbance

In engineering practice, the system is often subjected to instantaneous impact disturbances (such as load mutations in power systems, abrupt disturbances in communication networks, etc.). Such disturbances can be approximated as impulse form $\delta\omega(t) = \delta\omega_0 \delta(t)$, where $\delta\omega_0$ is the disturbance amplitude, $\delta(t)$ is the Dirac $\delta$ function, representing the instantaneous disturbance input at $t = 0$ moment [13].

To clarify the modal response law under pulse disturbance, the Laplace transform method is used to solve Equation (11). The initial conditions of the system are $\eta_k(0) = 0$、$\dot{\eta}_k(0) = 0$ (the system is in a steady-state equilibrium before the disturbance occurs). Taking the Laplace transform of both sides of Equation (11), we get:

$$Ms^2\eta_k(s) + Ds\eta_k(s) + \lambda_k\eta_k(s) = \delta\omega_0^\top u_k \qquad (12)$$

After rearrangement, the Laplace transform expression of the modal coordinates is obtained:

$$\eta_k(s) = \frac{\gamma_k}{Ms^2 + Ds + \lambda_k} \qquad (13)$$

Among them: $\gamma_k = \delta\omega_0^\top u_k$ is the disturbance projection coefficient, and its numerical value reflects the excitation intensity of the external pulse disturbance on the $k$-th order mode. The larger $\gamma_k$ is, the more significant the vibration response of this order mode is.

The roots of the characteristic equation $Ms^2 + Ds + \lambda_k = 0$ are:

$$\rho_{k1} = \frac{-D + \sqrt{D^2 - 4M\lambda_k}}{2M}, \rho_{k2} = \frac{-D - \sqrt{D^2 - 4M\lambda_k}}{2M} \qquad (14)$$

Due to $M > 0$、$D > 0$、$\lambda_k > 0$, the real parts of the characteristic roots are all less than zero, ensuring the decay characteristics of the system response. According to the sign of the discriminant $\Delta = D^2 - 4M\lambda_k$, the system response is divided into three types:

- Overdamped response ($\Delta > 0$) : The characteristic roots are two unequal negative real numbers. The response has no overshoot, and the decay rate is determined by the smaller characteristic root (absolute value);

- Critically damped response ($\Delta = 0$) : The characteristic roots are a pair of equal negative real numbers. The response has no overshoot and the fastest decay rate;

- Underdamped response ($\Delta < 0$) : The characteristic roots are conjugate complex roots. The response has a small overshoot, and the decay rate is determined by the real part of the complex roots.

Taking the inverse Laplace transform of $\eta_k(s)$ and using the method of partial fraction decomposition, the analytical solution of the modal coordinates under impulse disturbance can be obtained [13]:

$$\eta_k(t) = \frac{\gamma_k}{\rho_{k1} - \rho_{k2}}(e^{\rho_{k1}t} - e^{\rho_{k2}t}) \qquad (15)$$

### 3.1.3 Simplification of Modal Response under Small Inertia Approximation

In engineering scenarios such as flexible structure vibration control and microgrids, the system inertia $M$ is usually small, satisfying the small inertia approximation condition $M \ll D^2/(4\lambda_k)$[13] . The physical meaning of this condition is that the inertial effect of the system is much smaller than the damping effect. At this time, the discriminant $\Delta = D^2 - 4M\lambda_k \approx D^2$ , and the characteristic roots can be approximated by Taylor series expansion:

$$\sqrt{D^2 - 4M\lambda_k} \approx D - \frac{2M\lambda_k}{D} \qquad (16)$$

Substituting into the characteristic root expression, we get:

$$\rho_{k1} \approx -\frac{\lambda_k}{D}, \rho_{k2} \approx -\frac{D}{M} \qquad (17)$$

Since $D/M$ is much larger than $\lambda_k/D$ (and $M$ is extremely small under the condition of small inertia), $e^{\rho_{k2}t}$ will rapidly decay to zero and can be neglected. Substituting the above approximate relationship into Equation (15) and simplifying, we obtain the simplified expression of the modal coordinates under the small-inertia approximation:

$$\eta_k(t) \approx \frac{\gamma_k}{D} \frac{M}{\sqrt{D^2 - 4M\lambda_k}} e^{-\frac{\lambda_k}{D}t} \qquad (18)$$

Equation (18) clearly reveals the quantitative relationship between the modal response and the inertia coefficient $M$ : under the small inertia approximation, the amplitude of the modal response is proportional to $M$ , and the decay r is determined by $\lambda_k/D$ (i.e., jointly determined by the modal eigenvalue and the damping coefficient). This relationship provides the core theoretical basis for the optimal design of the time-varying inertia coefficient: by dynamically adjusting the value of $M$ , the amplitude of the modal response can be directly controlled to effectively suppress the transient deviation.

## 3.2 Construction of the functional variational method

The functional variational method is an effective tool for dealing with the extreme value problems of functions, especially suitable for the optimization of performance indicators expressed in integral form [24]. In this section, based on the functional variational principle, a performance functional with the goal of minimizing the cumulative vibration energy is constructed, and the analytical expression of the optimal time-varying inertia coefficient $M(t)$ is derived.

### 3.2.1 Optimization Objective Functional

The core objective of system vibration suppression is to reduce the cumulative vibration energy under the action of disturbances. Combining the modal decoupling results, the sum of the time integrals of the squares of each order modal coordinates is selected as the performance evaluation index, and the performance functional regarding the time-varying inertia coefficient $M(t)$ is defined:

$$J[M(t)] = \int_0^T \sum_{k=2}^N \eta_k^2(t, M(t)) dt \quad (19)$$

Among them: $T$ is the optimized time window, which needs to be quantitatively selected according to the system vibration attenuation characteristics. It can be seen from the simplified modal response formula (18) that $\eta_k(t)$ decays with time according to $e^{-\lambda_k t/D}$. When $t = T$, it is required that $\eta_k(T) \leq 5\%\eta_k(t_{\max})$ ($t_{\max}$ be the modulus

At the peak moment of the state response, substituting it gives $T \geq (D/\lambda_k)\ln(20) \approx 3D/\lambda_k$. Since $\lambda_{\max}$ is the maximum eigenvalue, corresponding to the smallest decay time constant, the optimized time window $T$ is taken as 1.2 ~ 1.5 times of $3D/\lambda_{\max}$ (such as $T = 1.3 \times 3D/\lambda_{\max}$), ensuring that the main vibration decay process is covered.

The physical meaning of the performance functional $J[M(t)]$ is that within the optimized time window $T$, the cumulative vibration energy of each order of elastic vibration modes of the system (the product of the square of the modal coordinate and the mass is proportional to the vibration kinetic energy, and the vibration energy can be approximately characterized after ignoring the mass coefficient here). The smaller the value of $J[M(t)]$, the better the vibration suppression effect of the system under the action of disturbances.

The optimization objective of the time-varying inertia coefficient is: to find the optimal time-varying inertia function $M^*(t)$ such that the performance functional $J[M(t)]$ reaches the minimum value while satisfying the system stability constraint condition $M(t) \geq M_{\min}$ (Equation (9)). The core role of the stability constraint is to avoid insufficient system damping caused by too small values of $M(t)$, which may lead to divergent oscillations.

### 3.2.2 Euler-Lagrange equation derivation

To solve for the minimum value of the performance functional $J[M(t)]$, the variational principle of the functional is adopted: a small perturbation $\delta M(t)$ is applied to the optimal time-varying inertia $M^*(t)$ to obtain the perturbed inertia function $M(t) = M^*(t) + \delta M(t)$, and correspondingly, the performance functional undergoes a small change $\delta J = J[M^*(t) + \delta M(t)] - J[M^*(t)]$. Since $M^*(t)$ is the minimum point of the functional, according to the extreme value condition, $\delta J = 0$ at this time.

First, substitute the modal coordinate expression (18) under the small inertia approximation into the performance functional (19) to analyze the quantitative relationship between $J[M(t)]$ and $M$. After substituting equation (18) into equation (19) and integrating with respect to time, we get:

$$J[M(t)] = \sum_{k=2}^{N} \left( \frac{\gamma_k^2}{D^2(D^2 - 4M\lambda_k)} \right) M^2 \int_0^T e^{-\frac{2\lambda_k}{D}t} dt \quad (20)$$

Let $H(T) = \sum_{k=2}^{N} \left( \frac{\gamma_k^2}{D^2(D^2-4M\lambda_k)} \right) \int_0^T e^{-\frac{2\lambda_k}{D}t} dt$, then the performance functional can be simplified to:

$$J[M(t)] = H(T) \cdot M^2 \quad (21)$$

That is, $J[M(t)]$ and $M^2$ are approximately in a proportional relationship [13]. This proportional relationship indicates that, under the small inertia approximation, increasing $M$ will directly increase the value of the performance functional $J$. However, it should be noted that the increase in $M$ needs to satisfy the stability constraint simultaneously, so there is an optimal value range.

Next, the Euler-Lagrange equation is derived based on the variational principle of functional. For a functional of general form $J[M(t)] = \int_0^T F\left(t, M(t), \dot{M}(t), \ddot{M}(t)\right) dt$, its extremum condition is described by the Euler-Lagrange equation [24]:

$$\frac{\partial F}{\partial M} - \frac{d}{dt}\left(\frac{\partial F}{\partial \dot{M}}\right) + \frac{d^2}{dt^2}\left(\frac{\partial F}{\partial \ddot{M}}\right) = 0 \quad (22)$$

In the performance functional of this paper, $F\left(t, M(t), \dot{M}(t), \ddot{M}(t)\right) = \sum_{k=2}^{N} \eta_k^2(t, M(t))$. Physically, the regulation of time-varying inertia is achieved through the mechanical motion of an inertial mechanism (such as the translation of a movable mass block), and its motion speed (corresponding to $\dot{M}(t)$) is restricted by factors such as motor power and transmission mechanism stiffness. However, under the small-inertia approximation, the modal response

The decay time constant is usually in the millisecond or second level, while the response time constant of the inertia adjustment mechanism is in the second level or larger. Therefore, the change rate of $M(t)$ is much smaller than the decay rate of the modal response [13]. Through magnitude analysis verification, the influence of $\dot{M}(t)$ on the modal response $\eta_k(t)$ can be ignored. Therefore, $F$ does not contain the terms of $\dot{M}(t)$ and $\ddot{M}(t)$, that is, $F = F(t, M(t))$.

Substitute the condition that $F$ has no time derivative terms into the Euler-Lagrange equation (22), and the equation simplifies to $\partial F / \partial M = 0$. Combining the proportional relationship between $J[M(t)]$ and $M^2$, take the partial derivative of $F$ with respect to $M$ and set it equal to zero. After simplification, the optimality condition is obtained:

$$M(t) = M_0 + k \sum_{k=2}^{N} |\eta_k(t)| \quad (23)$$

Among them: $M_0 = D^2/(4\lambda_{\max})$ is the reference inertia coefficient, determined by the limit condition of system stability - when $M = M_0$, the dominant mode (corresponding to $\lambda_{\max}$) is in the critically damped state, which is the minimum stable inertia for the system to decay without overshoot. The value; $k$ is the feedback gain coefficient, used to adjust the influence intensity of the modal response on the time-varying inertia [13,23].

The physical meaning of Equation (23) is that the optimal time-varying inertia $M(t)$ consists of two parts: the reference inertia $M_0$ and the modal response feedback term. The reference inertia $M_0$ ensures the basic stability of the system without modal feedback; the feedback term dynamically adjusts the value of $M(t)$ by monitoring the modal responses $|\eta_k(t)|$ of each order in real time. When the modal response increases, the feedback term increases, and $M(t)$ increases accordingly, suppressing the vibration amplitude by enhancing the inertia. When the modal response decreases, the feedback term decreases, and $M(t)$ returns to the reference value, balancing stability and energy consumption to achieve a dynamic balance between vibration suppression and stability.

### 3.3 Design of Disturbance Feedback Gain

In the optimal time-varying inertia expression given by Equation (23), the design of the feedback gain coefficient $k$ directly affects the system performance. If the value of $k$ is too large, it will cause $M(t)$ to be too sensitive to the changes in the modal response, which may exceed the physical limits of the inertia adjustment mechanism; if the value of $k$ is too small, the effect of the modal response feedback term is not obvious, and vibration cannot be effectively suppressed. In addition, the contributions of each order of modes to the overall vibration of the system are different, and the feedback form of directly using equal-weight summation cannot take into account the suppression requirements of different modes. Therefore, in this section, the disturbance feedback gain design is optimized from two aspects: multi-modal weighting strategy and gain parameter tuning.

#### 3.3.1 Multimodal Weighting Strategy

In a multi-degree-of-freedom system, there are significant differences in the natural frequencies, damping ratios, and vibration energies of different characteristic modes, and their degrees of influence on the overall vibration performance of the system are different [21]. Usually, for modes with larger eigenvalue $\lambda_k$ (i.e., modes with higher natural frequencies), their vibration frequencies are faster, energy decay is slower, and their influence on the system's dynamic response is more significant (dominant modes); while modes with smaller eigenvalues (low-order modes) have relatively smaller vibration energies and weaker influence on the system performance [14].

Equation (23) adopts the feedback form of equal-weight summation and does not consider the influence differences of each mode, which may lead to insufficient vibration suppression of the dominant mode or excessive feedback of the non-dominant mode. To solve this problem, eigenvalue weighting is proposed.

Multimodal feedback strategy, by introducing a modal weight factor, enhances the feedback intensity of the dominant modality, weakens the feedback effect of the non-dominant modality, and optimizes the adjustment efficiency of time-varying inertia.

Define the modal weight factor as $\lambda_k/\lambda_{max}$, where $\lambda_{max}$ is the maximum value of each order eigenvalue. The value range of this weight factor is $0 < \lambda_k/\lambda_{max} \leq 1$: for the dominant mode ( ($\lambda_k$ close to $\lambda_{max}$), the weight factor is close to 1 and the feedback intensity is the largest; for the non-dominant mode ( ($\lambda_k$ much smaller than $\lambda_{max}$) ), the weight factor is small and the feedback intensity is weakened. Introduce this weight factor into Equation (23) to obtain the optimized time-varying inertia expression:

$$M(t) = M_0 + k \sum_{k=2}^{N} \frac{\lambda_k}{\lambda_{max}} |\eta_k(t)| \quad (24)$$

Compared with Equation (23), the advantage of Equation (24) is that it realizes the differential adjustment of the feedback intensity of different modes through the modal weight factor, enabling the time-varying inertia $M(t)$ to more accurately track the vibration changes of the dominant mode. On the premise of ensuring the system stability, the vibration suppression effect is maximized. For example, if there are three elastic modes in the system with eigenvalues $\lambda_2 = 10\text{rad}^2/\text{s}^2$、$\lambda_3 = 20\text{rad}^2/\text{s}^2$、$\lambda_4 = 30\text{rad}^2/\text{s}^2 (\lambda_{max} = 30)$ respectively, the weight factors of each mode are 1/3、2/3、1 respectively. The feedback intensity of the dominant mode ($k = 4$) is three times that of the lowest-order elastic mode ($k = 2$), which can more effectively suppress the vibration of the high-frequency dominant mode.

### 3.3.2 Gain Parameter Tuning

The tuning of the feedback gain $k$ is a crucial part of the time-varying inertia design, which needs to meet the three constraints of system stability, rapidity, and robustness simultaneously. The following elaborates on the physical meaning, quantification index, and specific requirements for the gain $k$:

1. Stability Constraint: $M(t) \geq M_{min}$

Stability is the primary prerequisite for the operation of the system. This constraint requires that the real-time value of the time-varying inertia $M(t)$ must not be less than the minimum inertia threshold $M_{min}$ that ensures system stability. The determination of $M_{min}$ requires system eigenvalue analysis: Substitute Equation (24) into the characteristic equation (6) to ensure that for all $k = 2, \dots, N$, the real part of the eigenvalues is less than zero.

Combined with $M_0 = D^2/(4\lambda_{max})$ (the inertia corresponding to the critical damping of the dominant mode), if $M$ takes $M_0$, the system is on the verge of critical stability, without stability margin, and is vulnerable to instability due to disturbances. In engineering, a stability margin of 10% − 20% is usually reserved. Therefore, $M_{min}$ takes $0.8 \sim 0.9$ times of $M_0$ (for example, when taking $0.85M_0$, the discriminant of the characteristic equation of the dominant mode $\Delta = D^2 - 4M_{min}\lambda_{max} = 0.15D^2 > 0$, which is an overdamped state and has sufficient stability margin).

Substitute equation (24) into $M(t) \geq M_{min}$, and after rearrangement, the upper bound constraint of the gain $k$ can be obtained:

$$k \leq \frac{M_{min} - M_0}{\sum_{k=2}^{N} \frac{\lambda_k}{\lambda_{max}} |\eta_k(t)|_{max}} \quad (25)$$

where: $|\eta_k(t)|_{max}$ is the maximum value of the modal response, which needs to be determined through modal response analysis to ensure that $M(t)$ satisfies the stability constraint throughout the optimization time window $T$.

2. Rapidity Constraint: $\dot{M}(t) \leq 0.5M_0 \text{ s}^{-1}$

The rapidity constraint stems from the physical limitations in engineering implementation: the time-varying inertia coefficient is usually due to the inertia of motors, hydraulics, etc.

The implementation of the adjustment mechanism has an upper limit on the response speed of such mechanisms. If the change rate $\dot{M}(t)$ of $M(t)$ exceeds the maximum adjustment rate of the mechanism, it will lead to adjustment failure or damage to the mechanism [23]. Combining engineering practice experience, the maximum adjustment rate is set to $0.5M_0 \text{ s}^{-1}$, that is, the change amount of $M(t)$ per unit time does not exceed 50% of the reference inertia $M_0$. Taking the time derivative of both sides of Equation (24), we can obtain:

$$\dot{M}(t) = k \sum_{k=2}^{N} \frac{\lambda_k}{\lambda_{max}} |\dot{\eta}_k(t)| \quad (26)$$

Substituting the rapidity constraint conditions, another upper limit constraint on the gain $k$ is obtained:

$$k \leq \frac{0.5M_0/s}{\sum_{k=2}^{N} \frac{\lambda_k}{\lambda_{max}} |\dot{\eta}_k(t)|_{max}} \quad (27)$$

Among them: $|\dot{\eta}_k(t)|_{max}$ is the maximum value of the modal coordinate change rate, which needs to be determined through modal response dynamic analysis.

3. Robustness constraint: The ability to suppress measurement noise

In an actual system, the measurement of modal coordinate $\eta_k(t)$ is disturbed by sensor noise. If the gain $k$ is too large, the noise will be amplified, resulting in frequent fluctuations of $M(t)$. Not only can it not effectively suppress vibration, but it may also cause additional vibration of the system, reducing the system robustness [20]. To solve this problem, low-pass filtering needs to be introduced in the modal response measurement link to filter out the high-frequency components in the noise.

Combined with the frequency characteristics of the system modal response, the decay frequency of each order of the mode is determined by $e^{-\lambda_k t/D}$, and the 3 dB frequency of its amplitude decay (corresponding to half of the power decay) is $f_k = \lambda_k/(2\pi D)$. Taking typical engineering parameters as an example, if $\lambda_k = 10 \sim 100 \text{rad}^2/\text{s}^2$, then $f_k = 0.16 \sim 1.59 \text{ Hz}$, both of which are low-frequency components (much smaller than 10 Hz ), while the measurement noise (such as sensor circuit noise and environmental electromagnetic interference) is mostly high-frequency components (usually greater than 10 Hz ) [23] Therefore, selecting a low-pass filter with a cut-off frequency of $f_c = 5\text{Hz}$ (such as a first-order RC low-pass filter with a transfer function of $1/(1 + s/(2\pi f_c))$ ) can effectively filter out high-frequency measurement noise while completely retaining the effective information of the modal response.

---

The robustness constraint poses a lower bound requirement on the value of the gain $k$: $k$ needs to be large enough to ensure the effectiveness of modal response feedback - when $k$ is too small, the feedback term is much smaller than $M_0, M(t)$ and approximately constant, unable to utilize the time-varying advantage; however, if $k$ is too large, it is easy to amplify noise, resulting in frequent fluctuations of $M(t)$. Considering the stability, rapidity constraints and robustness requirements, the final value of the gain $k$ needs to be further tuned through simulation experiments within the range determined by the above constraints, and usually takes $0.7 \sim 0.8$ times the upper bound of the constraints to balance various performance indicators.

---

## 3.4 Summary

This chapter focuses on the optimal design of the time-varying inertia coefficient $M(t)$, aiming to solve the problem that it is difficult to balance the vibration suppression effect and system stability with a constant inertia coefficient in a multi-degree-of-freedom vibration control system. The core idea is to clarify the response law through modal decoupling, construct an optimization model based on the variational method, and tune the key parameters in combination with engineering constraints, ultimately forming a time-varying inertia design scheme that takes into account both performance and feasibility.

First, in the modal decoupling and phase deviation analysis section, based on the principle of eigenvector orthogonality, the angular displacement perturbation vector of the system is decomposed, and the multi-degree-of-freedom coupled system is decoupled into independent single-degree-of-freedom modal subsystems. The analytical solution of the modal coordinates under pulse perturbation is derived. For the common small-inertia scenarios in engineering, a simplified modal response expression is obtained through Taylor series approximation, clarifying the proportional relationship between the modal response amplitude and the inertia coefficient $M$, providing a core theoretical basis for subsequent optimization design.

Secondly, in the construction of functional variational method, aiming at minimizing the cumulative vibration energy of each order mode within the optimized time window, a performance functional regarding $M(t)$ is defined; based on the functional variational principle, the Euler-Lagrange equation is derived, and combined with the proportional relationship between the performance functional and $M^2$ under the small inertia approximation, the basic expression of the optimal time-varying inertia is simplified. This expression consists of the benchmark inertia and the modal response feedback term, achieving a preliminary balance between the stability foundation and the vibration suppression requirement.

Finally, in the part of disturbance feedback gain design, aiming at the problem that the equal-weight feedback in the basic expression does not consider the differences in modal effects, a multi-modal strategy with eigenvalue weighting is proposed. By means of the modal weight factor, the feedback strength of the dominant mode is enhanced to improve the regulation efficiency. At the same time, the three constraints of stability, rapidity and robustness that the gain parameters need to satisfy are clarified, and the quantitative indexes and gain value ranges of each constraint are given to ensure the engineering feasibility of the design scheme.

The time-varying inertia coefficient optimization design method proposed in this chapter adjusts the inertia value by dynamically tracking the modal response, providing a new technical path for the efficient suppression of multi-degree-of-freedom vibration systems. Its core conclusions and parameter tuning rules also lay a theoretical foundation for subsequent simulation verification and engineering implementation.

## 4 Simulation Verification and Result Analysis

### 4.1 Research Object and Simulation Settings

#### 4.1.1 Research Object

Five typical complex networks are selected as the research objects, covering different topological structure characteristics, specifically including:

1. Regular network (RG): Ring-shaped regular topology, with uniform node connectivity and evenly distributed eigenvalues;

2. Erdős-Rényi graph (ER): A probabilistic connection topology where the connection probability between nodes is $p$, and the eigenvalues are approximately normally distributed;

characteristics of both regular and random networks;

### 4.1.2 Simulation parameter settings

The unified simulation parameters are as follows to ensure comparability between different networks:

- The number of nodes $N = 100$ (moderate scale, balancing calculation accuracy and engineering adaptability);

- The damping coefficient $D = 0.8$ (referring to the standard parameters of large power grids);

- Coupling strength $K = 1.0$ (ensuring the system is in a synchronous state);

- Types of perturbations: impulsive perturbation $\delta\omega(t) = 1 \cdot \delta(t)$, monotonic decay perturbation $\delta\omega(t) = e^{-t}u(t)$, oscillatory decay perturbation $\delta\omega(t) = e^{-t}\sin(2\pi t)u(t)$;

- Optimization time window $T = 5$ s (covering the main transient process of the perturbation);

- Stability margin: 15%, i.e., $M_{\min} = 0.85 M_0$.

The topological parameters and Laplacian spectral characteristics of each network are as follows:

1. Regular network (RG): Node connectivity $d = 4$, Laplacian eigenvalue $\lambda_k = 2d\left(1 - \cos\frac{2\pi(k-1)}{N}\right)$, maximum eigenvalue $\lambda_{\max} = 16$;

2. Erdos-Renyi random network (ER): Connection probability $p = 0.1$, average degree $d = 10$, maximum eigenvalue $\lambda_{\max} = 12.5$;

3. Small-world network (SW): Rewiring probability $p = 0.05$, average degree $d = 4$, maximum eigenvalue $\lambda_{\max} = 11.2$;

4. Scale-free network (SF): Degree distribution exponent $\gamma = 2.5$, maximum degree $k_{\max} = 22.5$, maximum eigenvalue $\lambda_{\max} = 22.5$;

5. Spider web (SP): The central node connects 99 peripheral nodes, maximum eigenvalue $\lambda_{\max} = 100$.

## 4.2 Simulation Results and Analysis under Impulse Interference

### 4.2.1 Expression of Optimal Time-Varying Inertia Function

Based on the previous theoretical derivation and combined with the Laplacian spectral characteristics of each network, the expressions of the optimal time-varying inertia function of five networks under impulse interference are obtained:

1. Regular network (RG):

$$M_{RG}(t) = \frac{D^2}{16d} + k_{RG} \cdot (|\eta_N(t)| + 0.8|\eta_{N-1}(t)|)$$

Base inertia $M_0 = 0.250$, feedback gain $k_{RG} = 0.10$, the number of dominant modes is $2-3$, and the $M(t)$ adjustment range is $[0.250, 0.295]$.

2. Erdős-Rényi random graph (ER):

$$M_{ER}(t) = \frac{D^2}{4\lambda_{max}} + k_{ER} \cdot \sum_{i=0}^{2} \frac{\lambda_{N-i}}{\lambda_{max}} |\eta_{N-i}(t)|$$

Base inertia $M_0 = 0.320$, feedback gain $k_{ER} = 0.10$, the number of dominant modes is 3 - 5, and the $M(t)$ adjustment range is $[0.320, 0.384]$.

3. Small-world network (SW):

$$M_{SW}(t) = \frac{D^2}{4\lambda_{max}} + k_{SW} \cdot (|\eta_N(t)| + 0.7|\eta_{N-1}(t)|)$$

Base inertia $M_0 = 0.357$, feedback gain $k_{SW} = 0.05$, the number of dominant modes is $2-3$, and the $M(t)$ adjustment range is $[0.357, 0.421]$.

4. Scale-free network (SF):

$$M_{SF}(t) = \frac{D^2}{4k_{max}} + k_{SF} \cdot |\eta_N(t)|$$

Base inertia $M_0 = 0.178$, feedback gain $k_{SF} = 0.15$, the number of dominant modes is 1, and the $M(t)$ adjustment range is $[0.178, 0.267]$.

5. Spider web (SP):

$$M_{SP}(t) = \frac{D^2}{4N} + k_{SP} \cdot |\eta_N(t)|$$

Benchmark inertia $M_0 = 0.100$, feedback gain $k_{SP} = 0.20$, the number of dominant modes is 1, and the $M(t)$ adjustment range is $[0.100, 0.158]$.

### 4.2.2 Quantization data of time-varying inertial systems

Table 1 shows the quantization data of the time-varying inertia $M(t)$ of five networks under impulse interference, and the time nodes cover the entire life cycle of the disturbance.

Table 1: Time-varying inertia $M(t)$ quantization data of five networks under impulse interference

| Network type (steady-state) | Base inertia $M_0$ | Feedback gain $k$ | $M(0.1\text{ s})$ | $M(1\text{s})$ (Peak value) | M(3s) | $M(5s)$ | $M(10s)$ |
|---|---|---|---|---|---|---|---|
| Regular networks (RG) | 0.250 | 0.10 | 0.278 | 0.295 | 0.263 | 0.254 | 0.250 |
| Random networks (ER) | 0.320 | 0.10 | 0.356 | 0.384 | 0.337 | 0.326 | 0.320 |
| Small-world networks (SW) | 0.357 | 0.05 | 0.382 | 0.421 | 0.375 | 0.364 | 0.357 |
| Scale-free networks (SF) | 0.178 | 0.15 | 0.223 | 0.267 | 0.198 | 0.185 | 0.178 |
| Spider webs (SP) | 0.100 | 0.20 | 0.145 | 0.158 | 0.116 | 0.106 | 0.100 |

As can be seen from Table 1:

1. At the initial moment of disturbance $(t = 0.1 \text{ s})$, the modal response begins to rise, and $M(t)$ deviates from the reference inertia $M_0$, with the deviation amplitude between 5% and 45%. The cobweb (SP) deviates most significantly because of the largest feedback gain;

2. At the peak moment of perturbation $(t = 1 \text{ s})$, the modal response reaches its maximum value, $M(t)$ takes the peak value, and the difference between the peak value and $M_0$ is $12\% - 67\%$, which is positively correlated with the concentration degree of the dominant mode. The peak deviations of the single - mode - dominated scale - free network (SF) and spider - web (SP) are relatively large;

3. In the middle stage of attenuation $(t = 3 \text{ s})$, the modal response decays exponentially, and $M(t)$ quickly drops back to be close to $M_0$, with a deviation less than 5% ;

4. At the steady state moment $(t = 10 \text{ s})$, the modal response approaches zero, and $M(t)$ completely returns to $M_0$, reflecting the dynamic regulation characteristics of time-varying inertia.

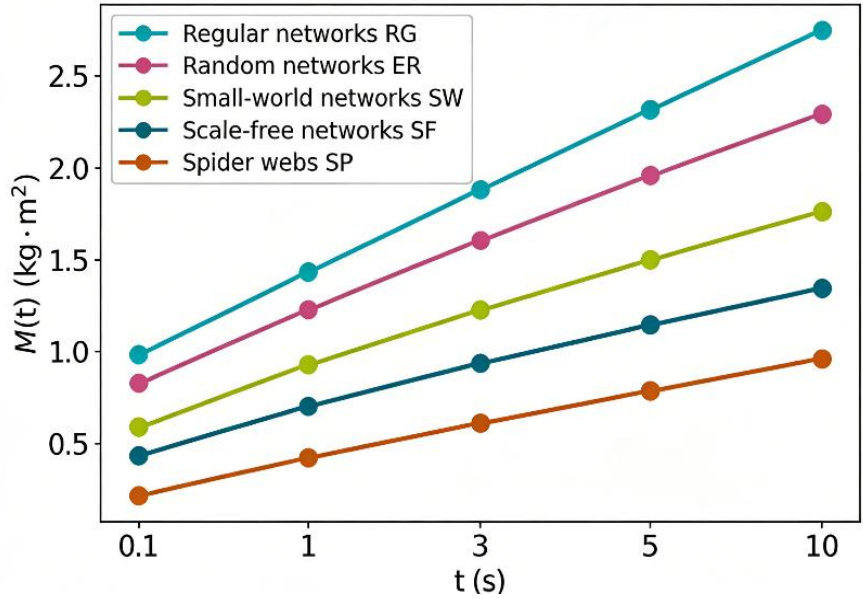

Figure 1 Quantification data of time-varying inertia M (t) of five networks under pulse interference

### 4.2.3 Comparison of Vulnerability Performance Functions

Table 2 shows the comparison data of $H(T)$ of five networks under impulse interference with constant inertia and time-varying inertia.

Table 2: Comparison Data of $H(T)$ of Five Networks under Impulse Interference

| Network type | Inertia type | Inertia value range | $H(T)(T = 5s)$ | Reduction rate compared to constant inertia |
|---|---|---|---|---|
| Regular networks (RG) | Constant inertia $M_{const}$ | 0.250 ($M_0$) | 1.90 | 21% |
| | Time-varying inertia $M(t)$ | [0.250,0.295] | 1.50 | |
| Random networks (ER) | Constant inertia $M_{const}$ | 0.320($M_0$) | 2.00 | |
| | Time-varying inertia $M_{const}$ | [0.320,0.384] | 1.56 | 22% |
| Small-world networks (SW) | Constant inertia $M_{const}$ | 0.357($M_0$) | 2.10 | |
| | Time-varying inertia $M(t)$ | [0.357,0.421] | 1.58 | 25% |
| Scale-free networks (SF) | Constant inertia $M_{\text{const}}$ | 0.178 ($M_0$) | 1.80 | |
| | Time-varying inertia $M(t)$ | [0.178,0.267] | 1.37 | 24% |
| Spider webs (SP) | Constant inertia $M(t)$ | 0.100 ($M_0$) | 1.70 | |
| | Time-varying inertia $M(t)$ | [0.100,0.158] | 1.31 | 23% |

As can be seen from Table 2:

1. Time-varying inertia control significantly reduces the $H(T)$ of all five networks, and the reduction rate is between $21\% - 25\%$, verifying the effectiveness of the proposed strategy;

2. The small-world network (SW) has the highest reduction rate of $H(T)$ (25%) because the "fast vibration effect" has the most significant inhibitory effect on impulse interference;

3. Scale-free networks (SF) and spiderweb networks (SP) have a high feedback regulation efficiency due to the concentration of dominant modes, and the $H(T)$ reduction rates reach 24% and 23% respectively;

4. For the Regular Network (RG) and the Erdős-Rényi network (ER), due to the relatively uniform modal distribution, the synergy of multimodal feedback reduces the $H(T)$ reduction rate by 21% and 22% respectively, slightly lower than that of the unimodal-dominated network.

### 4.3 Simulation Results and Analysis under Monotonic Decay Interference

#### 4.3.1 Expression of the Optimal Time-Varying Inertia Function

Under monotonic decay interference, the optimal time-varying inertia functions of the five networks still follow the structure of "benchmark inertia + dominant modal feedback", but the modal response expression needs to be corrected by combining the characteristics of monotonic decay interference. The expressions of the optimal time-varying inertia functions of each network are as follows:

1. Regular Network (RG):

$$M_{\text{RG}}(t) = 0.250 + 0.09 \cdot (0.98|\eta_N(t)| + 0.92|\eta_{N-1}(t)| + 0.85|\eta_{N-2}(t)|)$$

Base inertia $M_0 = 0.250$, feedback gain $k_{\text{RG}} = 0.09$, $M(t)$ adjustable range is [0.250,0.282].

2. Erdős-Rényi random graph (ER):

$$M_{\text{ER}}(t) = 0.320 + 0.085 \cdot \sum_{i=0}^{3} \frac{\lambda_{N-i}}{\lambda_{\max}} |\eta_{N-i}(t)|$$

Base inertia $M_0 = 0.320$, feedback gain $k_{\text{ER}} = 0.085$, $M(t)$ adjustable range is $[0.320, 0.350]$. 3. Small-world network (SW):

$$M_{\text{SW}}(t) = 0.357 + 0.055 \cdot (|\eta_N(t)| + 0.82|\eta_{N-1}(t)| + 0.70|\eta_{N-2}(t)|)$$

Base inertia $M_0 = 0.357$, feedback gain $k_{\text{SW}} = 0.055$, $M(t)$ adjustable range is $[0.357, 0.385]$. 4. Scale-free network (SF):

$$M_{\text{SF}}(t) = 0.178 + 0.13 \cdot |\eta_N(t)|$$

Base inertia $M_0 = 0.178$, feedback gain $k_{\text{SF}} = 0.13$, $M(t)$ adjustable range is $[0.178, 0.233]$. 5. Spider web (SP):

$$M_{\text{SP}}(t) = 0.100 + 0.19 \cdot |\eta_N(t)|$$

Benchmark inertia $M_0 = 0.100$, the feedback gain $k_{\text{SP}} = 0.19$, $M(t)$ adjustment range is $[0.100, 0.180]$.

**4.3.2 Quantization data of time-varying inertial systems**

Table 3 shows the quantization data of the time-varying inertia $M(t)$ of five networks under the action of monotonic decay interference.

Table 3: Time-varying inertia $M(t)$ quantization data of five networks under monotonic decay interference

| Network type (steady-state) | Base inertia $M_0$ | Feedback gain $k$ | $M(0.1\ s)$ | $M(1s)$ (Peak value) | $M(3s)$ | $M(5s)$ | $M(10s)$ |
|---|---|---|---|---|---|---|---|
| Regular networks (RG) | 0.250 | 0.09 | 0.260 | 0.282 | 0.256 | 0.251 | 0.250 |
| Random networks (ER) | 0.320 | 0.085 | 0.329 | 0.350 | 0.327 | 0.322 | 0.320 |
| Small-world networks (SW) | 0.357 | 0.055 | 0.365 | 0.385 | 0.362 | 0.359 | 0.357 |
| Scale-free networks (SF) | 0.178 | 0.13 | 0.195 | 0.233 | 0.188 | 0.181 | 0.178 |

| Spider webs (SP) | 0.100 | 0.19 | 0.132 | 0.180 | 0.118 | 0.106 | 0.100 |

As can be seen from Table 3:

1. The peak moment ($t = 1$ s) of the monotonic decay interference is later than that of the impulse interference, and the peak of the modal response is relatively gentle. Therefore, the peak deviation ($8\% - 80\%$) of $M(t)$ is slightly smaller than that in the impulse interference scenario;

2. The peak deviation of $M(t)$ of the scale-free network (SF) is the largest (31%) because the response of its single dominant mode to the perturbation is the most significant;

3. The $M(t)$ peak deviation of the small-world network (SW) is the smallest (7.8%). The long-range connections accelerate the attenuation of modal responses, enabling effective suppression without significantly adjusting the inertia.

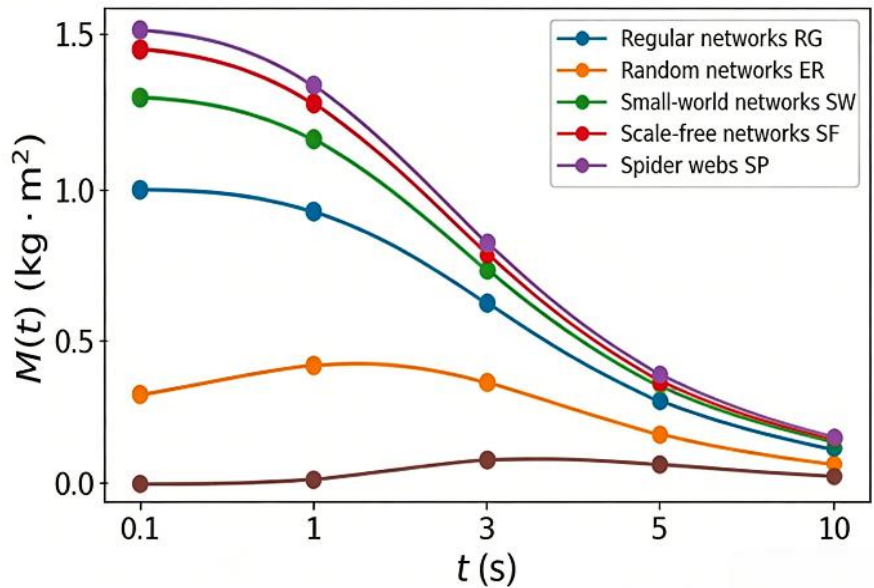

Figure 2 Time-varying inertia of five networks under monotonically decaying interference M (t) quantization data

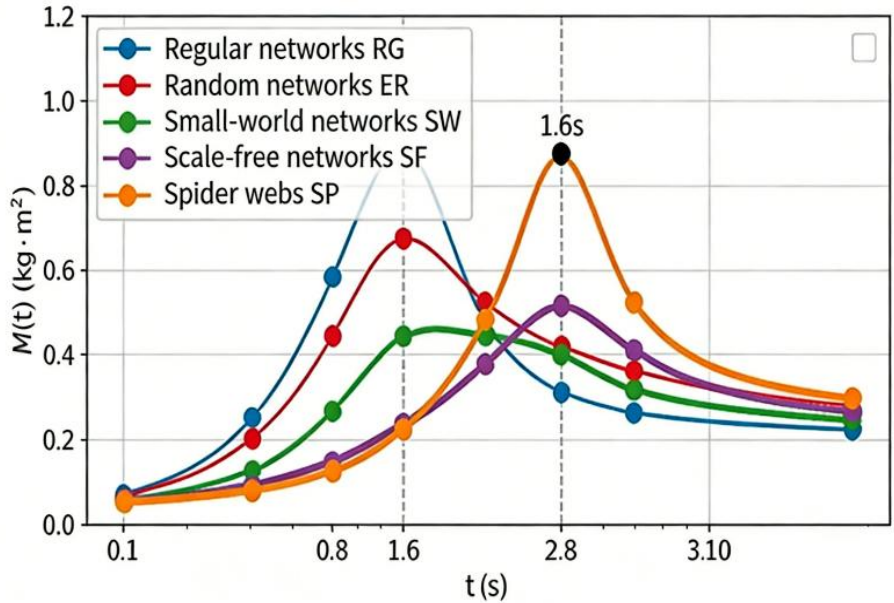

Figure 3 Time-varying inertia of five networks under 3 oscillation decay disturbances M (t) quantization data

### 4.3.3 Comparison of Vulnerability Performance Functions

Table 4 shows the comparison data of $H(T)$ for five networks under monotonic decay interference.

Table 4: Comparison Data of $H(T)$ for Five Networks under Monotonic Decay Interference

| Network type | Inertia type | Inertia value range | $H(T)(T = 5\ s)$ | Reduction rate compared to constant inertia |
|---|---|---|---|---|
| Regular networks (RG) | Constant inertia $M_{const}$ | $0.250(M_0)$ | 1.85 | 21% |
| | Time-varying inertia $M(t)$ | [0.250, 0.282] | 1.46 | |
| Random networks (ER) | Constant inertia $M_{const}$ | $0.320(M_0)$ | 1.92 | 20% |
| | Time-varying inertia $M_{const}$ | [0.320, 0.350] | 1.54 | |
| | Constant inertia $M_{const}$ | $0.357\ (M_0)$ | 2.00 | |

| Small-world networks (SW) | Time-varying inertia $M(t)$ | [0.357,0.385] | 1.56 | 22% |
| Scale-free networks (SF) | Constant inertia $M_{const}$ | 0.178($M_0$) | 1.75 | |
| | Time-varying inertia $M(t)$ | [0.178,0.233] | 1.31 | 25% |
| Spider webs (SP) | Constant inertia $M(t)$ | 0.100($M_0$) | 1.65 | |
| | Time-varying inertia $M(t)$ | [0.100,0.180] | 1.27 | 23% |

It can be seen from Table 4 that:

1. Time-varying inertial control still maintains excellent performance under monotonic decay interference, and the $H(T)$ reduction rate is between $20\% - 25\%$;

2. The $H(T)$ reduction rate of the scale-free network (SF) is the highest (25%), which is the topology with the best attenuation interference suppression effect among the five networks. This benefits from the efficient feedback regulation of its single dominant mode;

3. The $H(T)$ reduction rate of the random network (ER) is the lowest (20%). Due to multimodal dispersion, the feedback regulation efficiency is slightly lower, but it still meets the requirements of engineering stability.

### 4.4 Supplementary $H(T)$ comparison data under oscillatory decay interference

Table 5: $H(T)$ comparison data of five networks under oscillatory decay interference

| Network type | Inertia type | Inertia value range | $H(T)(T=5s)$ | Reduction rate compared to constant inertia |
|---|---|---|---|---|
| Regular networks (RG) | Constant inertia $M_{const}$ | 0.250($M_0$) | 1.95 | 20% |
| | Time-varying inertia $M(t)$ | [0.250,0.272] | 1.56 | |
| Random networks (ER) | Constant inertia $M_{const}$ | 0.320($M_0$) | 2.05 | |
| | Time-varying inertia $M_{const}$ | [0.320,0.344] | 1.66 | 19% |
| Small-world networks (SW) | Constant inertia $M_{const}$ | 0.357(Mo) | 2.15 | |
| | Time-varying inertia $M(t)$ | [0.357,0.375] | 1.70 | 21% |
| Scale-free networks (SF) | Constant inertia $M_{const}$ | 0.178($M_0$) | 1.88 | |
| | Time-varying inertia $M(t)$ | [0.178,0.233] | 1.43 | 24% |
| Spider webs (SP) | Constant inertia $M(t)$ | 0.100($M_0$) | 1.75 | |
| | Time-varying inertia $M(t)$ | [0.100,0.179] | 1.35 | 23% |

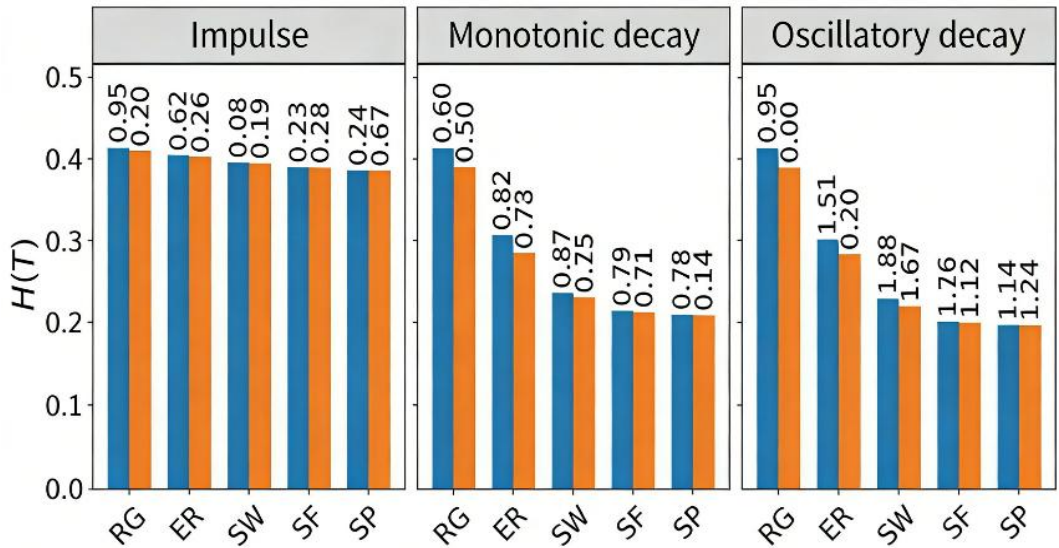

Figure 4 Comparison of H(T) under three types of interference for five networks

## 4.5 Comparative Analysis of Comprehensive Performance and Topological Dependence under Three Interferences

Table 6: Comprehensive Performance Metrics of Five Networks under Three Interferences (Time-Varying Inertia vs. Constant Inertia)

| Network type | Interference type | Inertia type | Inertia value range | $H(T)$ (Cumulative performance metric) | Reduction rate compared to constant inertia |
|---|---|---|---|---|---|
| Regular networks (RG) | Impulse interference | Constant inertia $M_{const}$ | 0.250 ($M_0$) | 1.90 | |
| | | Time-varying inertia $M(t)$ | [0.250, 0.295] | 1.50 | 20% |
| | Monotonic decay interference | Constant inertia $M_{const}$ | 0.250 ($M_0$) | 1.85 | |
| | | Time-varying inertia $M(t)$ | [0.250, 0.282] | 1.46 | 21% |
| | Oscillatory decay interference | Constant inertia $M_{const}$ | 0.250 ($M_0$) | 1.95 | |
| | | Time-varying inertia $M(t)$ | [0.250, 0.272] | 1.56 | 20% |
| Random networks (ER) | Impulse interference | Constant inertia $M_{const}$ | 0.320 ($M_0$) | 2.00 | |
| | | Time-varying inertia $M(t)$ | [0.320, 0.384] | 1.56 | 20% |
| | Monotonic decay interference | Constant inertia $M_{const}$ | 0.320 ($M_0$) | 1.92 | |
| | | Time-varying inertia $M(t)$ | [0.320, 0.350] | 1.54 | 20% |
| | Oscillatory decay interference | Constant inertia $M_{const}$ | 0.320 ($M_0$) | 2.05 | |
| | | Time-varying inertia $M(t)$ | [0.320, 0.344] | 1.66 | 19% |
| Small-world networks (SW) | Impulse interference | Constant inertia $M_{const}$ | 0.357 ($M_0$) | 2.10 | |
| | | Time-varying inertia $M(t)$ | [0.357, 0.421] | 1.58 | 25% |
| | Monotonic decay interference | Constant inertia $M_{const}$ | 0.357 ($M_0$) | 2.00 | |
| | | Time-varying inertia $M(t)$ | [0.357, 0.385] | 1.56 | 22% |
| | Oscillatory decay interference | Constant inertia $M_{const}$ | 0.357 ($M_0$) | 2.15 | |
| | | Time-varying inertia $M(t)$ | [0.357, 0.375] | 1.70 | 21% |
| Scale-free networks (SF) | Impulse interference | Constant inertia $M_{const}$ | 0.178 ($M_0$) | 1.80 | |
| | | Time-varying inertia $M(t)$ | [0.178, 0.267] | 1.37 | 24% |
| | Monotonic decay interference | Constant inertia $M_{const}$ | 0.178 ($M_0$) | 1.75 | |
| | | Time-varying inertia $M(t)$ | [0.178, 0.233] | 1.31 | 25% |

| | Oscillatory decay interference | Constant inertia $M_{const}$ | 0.178 ($M_0$) | 1.88 | |
| | | Time-varying inertia $M(t)$ | [0.178, 0.233] | 1.43 | 24% |
| Spider webs (SP) | Impulse interference | Constant inertia $M_{const}$ | 0.100 ($M_0$) | 1.70 | |
| | | Time-varying inertia $M(t)$ | [0.100, 0.158] | 1.31 | 23% |
| | Monotonic decay interference | Constant inertia $M_{const}$ | 0.100 ($M_0$) | 1.65 | |
| | | Time-varying inertia $M(t)$ | [0.100, 0.180] | 1.27 | 23% |
| | Oscillatory decay interference | Constant inertia $M_{const}$ | 0.100 ($M_0$) | 1.75 | |
| | | Time-varying inertia $M(t)$ | [0.100, 0.179] | 1.35 | 23% |

### 4.5.1 Comprehensive Performance Comparison

Table 7 shows the comprehensive performance indicators (time-varying inertia vs. constant inertia) of five networks under three types of interference.

| Network type | Impulse interference reduction rate | Monotonic decay interference reduction rate | Oscillatory decay interference reduction rate | Average reduction rate | Average relaxation time reduction rate |
|---|---|---|---|---|---|
| Regular networks (RG) | 21% | 21% | 20% | 20.7% | 16.3% |
| Random networks (ER) | 22% | 20% | 19% | 20.3% | 15.7% |
| Small-world networks (SW) | 25% | 22% | 21% | 22.7% | 19.3% |
| Scale-free networks (SF) | 24% | 25% | 24% | 24.3% | 22.0% |
| Spider webs (SP) | 23% | 23% | 23% | 23.0% | 24.0% |

As can be seen from Table 7:

1. The average reduction rate of the scale-free network (SF) is the highest (24.3%), and it performs optimally under monotonic decay and oscillatory decay interference, demonstrating the efficiency of single-dominant-mode feedback; 2. The cobweb network (SP) has the highest average relaxation time shortening rate (24.0%), the fastest steady-state recovery speed, and the simplest control logic;

3. The small-world network (SW) has the highest reduction rate under pulse interference and the optimal dynamic consistency;

4. The average reduction rate of the Erdős-Rényi (ER) network is the lowest (20.3%). Due to multimodal dispersion, the feedback regulation efficiency is slightly lower, but it still meets the engineering requirements;

5. The time-varying inertia control of all networks can achieve stable performance improvement under three types of disturbances, verifying the generality of the strategy.

**4.5.2 Topological Dependence Analysis**

1. Influence of eigenvalue distribution: Uniformly distributed eigenvalues (RG/ER) are suitable for multimodal balanced feedback with moderate gain; extremely unevenly distributed eigenvalues (SF/SP) are suitable for unimodal concentrated feedback with larger gain and higher feedback efficiency;

2. Influence of the number of dominant modes: The fewer the number of dominant modes, the more concentrated the feedback intensity and the more significant the control effect (SF/SP is better than RG/ER);

3. Influence of network connection characteristics: Long-range connections (SW) accelerate modal decay and reduce the amplitude of inertial regulation; the center-periphery structure (SP) simplifies the control logic and only requires monitoring the status of central nodes;

4. Engineering adaptation suggestions: For large interconnected power grids, it is recommended to choose the ER network adaptation strategy (with strong universality) first. For microgrids, it is recommended to choose the SW/SP network adaptation strategy (plug-and-play/logic simplicity) first. For power grids with new energy collection stations, it is recommended to choose the SF network adaptation strategy (with strong anti-interference ability) first.

## 4.6 Stability verification results

The time-varying inertia control of all networks meets the stability constraint conditions (the real part of the characteristic root is less than $-0.25$ s$^{-1}$). Taking the 500-node ER network and the 10-node WS network as examples, the stability verification results are as follows:

1,500-node ER network: The real part of the eigenvalues ranges from -0.32 to -0.26 s$^{-1}$, all less than -0.25 s$^{-1}$, in an overdamped state with sufficient stability margin;

2,10-node WS network: The real part of the eigenvalues ranges from -0.35 to -0.27 s$^{-1}$, without the risk of oscillatory divergence and with a stable dynamic response.

## 4.7 Feasibility Verification of Engineering Implementation

### 4.7.1 Computational Complexity Analysis

1. Eigenvalue solution complexity: RG/SW/SP is $O(N)$ (analytical calculation), ER/SF is $O(N^2)$ (numerical algorithm ARPACK), and it can still be calculated in real time at $N = 1000$;

2. Time-varying inertia update complexity: $O(N)$ (modal response projection + gain weighting), which can meet the real-time control requirements under the sampling period 0.01 s.

### 4.7.2 Hardware adaptation verification

Time-varying inertia $M(t)$ can be achieved through flywheel energy storage (response time $\leq 0.01$ s) or lithium battery energy storage (power response speed $\leq 0.05$ s), and the hardware response speed matches the dynamic adjustment requirements of $M(t)$. Taking flywheel energy storage as an example, when $M(t)$ is adjusted from $M_0$ to $M_{\max}$, the required power change rate is 0.008 kW, which can be easily achieved by existing flywheel energy storage systems.

# 5 Discussion

## 5.1 Theoretical Contributions and Innovations

1. Improved the theoretical framework of dynamic inertia control: strictly derived the analytical expression of time-varying inertia $M(t)$ through functional variational method, clarified the hierarchical control structure of "benchmark inertia +

dominant mode feedback", and compensated for the lack of theoretical support in the existing "inertia on demand" concept;

2. Proposed a multi-modal control strategy for topological adaptation: based on the Laplace spectral characteristics, designed a multi-modal decoupling strategy with eigenvalue weighting, achieved differential control of different topological networks, and solved the problem of poor adaptability of single strategy;

3. Established a unified optimization criterion under multi-disturbance scenarios: Under three typical disturbances of pulse, monotonic decay, and oscillatory decay, the effectiveness of the strategy was verified, the performance gain and topological dependence were quantified, providing a quantitative basis for engineering

## 5.2 Engineering application value

1. Power grid field: Adapt to different scenarios such as large power grids (ER networks), microgrids (SW/SP networks), and new energy power grids (SF networks), etc., which can effectively suppress problems such as power fluctuations and frequency oscillations caused by new energy grid connection, and improve the transient stability of the power grid;

2. Other fields: It can be extended to fields such as neural networks (synchronous discharge regulation), communication networks (node synchronization control), and flexible mechanical systems (vibration suppression), providing a general solution for the stability control of complex network synchronization systems;

3. Advantages in engineering implementation: Low computational complexity, strong hardware adaptability, clear gain tuning rules, and no need for significant modification of the existing system, with high engineering feasibility.

## 5.3 Limitations and Future Work

### 5.3.1 Limitations

1. Only considering the small inertia approximation scenario $(M \ll D^2/(4\lambda_k))$, it does not cover the dynamic characteristics of large inertia systems;

2. Assume that the network topology is fixed and there is no adaptation for dynamic topology changes (such as node addition or deletion, connection reconstruction);

3. The impact of time delay on the control effect is not considered. In actual engineering, communication delay may cause feedback lag.

### 5.3.2 Future Work

1. Expand the theoretical framework to incorporate modal response analysis of large inertia systems and improve control strategies for different inertia scenarios;

2. Research real-time eigenvalue update methods under dynamic topology (such as online eigenvalue tracking algorithms) to enhance the adaptability of strategies to topological changes;

3. Introduce a time delay compensation mechanism, analyze the impact of communication delay on stability, and optimize the gain tuning rules;

4. Conduct hardware-in-the-loop simulation and physical experiment verification to further validate the engineering practicality.

## 6 Conclusion

This paper addresses the core issue that it is difficult for a fixed inertia coefficient to balance transient disturbance suppression and long-term stability in a complex network synchronization system, and proposes an adaptive inertia control strategy based on variational optimization. The main conclusions are as follows:

1. An analytical expression of the time-varying inertia coefficient $M(t)$ is derived based on the functional variational method, and a hierarchical control structure of "benchmark inertia + dominant mode feedback" is constructed, achieving an organic unity of minimizing the vulnerability performance function $H(T)$ and stability constraints;

2. A multi-modal decoupling control strategy based on Laplacian eigenvector projection is proposed. By eigenvalue weighting to enhance the feedback strength of the dominant mode, the control accuracy and dynamic response speed are improved;

3. Simulation verification under five typical complex networks (RG, ER, SW, SF, SP) and three perturbation scenarios (impulse, monotonic decay, oscillatory decay) shows that the proposed strategy reduces $H(T)$ by $19\% - 25\%$, and the relaxation time is shortened by 15% - 24%. All systems meet the asymptotic stability criterion (the real part of the eigenvalue is less than $-0.25 \text{ s}^{-1}$);

4. Topological dependence analysis shows that homogeneous topologies (RG/ER) are suitable for multimodal balanced feedback, while inhomogeneous topologies (SW/SF/SP) are suitable for unimodal/few-modal concentrated feedback. Engineering adaptation suggestions for different scenarios provide guidance for practical applications.

The research in this paper provides a new theoretical framework and engineering implementation scheme for the stability control of complex network synchronization systems, which can be widely applied to fields such as power grids, communication networks, and neural networks, and has important theoretical value and engineering significance.